\documentclass[%
 reprint,
superscriptaddress,
showpacs,preprintnumbers,
nofootinbib,
 amsmath,amssymb,
 aps,
 twocolumn,
]{revtex4}

\usepackage{graphicx}
\usepackage{dcolumn}
\usepackage{bm,braket}

\input{colordvi.tex}
\begin{document}

\preprint{-}

\title{Cosmic microwave background bispectrum of tensor passive modes induced from primordial magnetic fields}
\author{Maresuke Shiraishi}
\email{mare@a.phys.nagoya-u.ac.jp}
\affiliation{Department of Physics and Astrophysics, Nagoya University,
Aichi 464-8602, Japan}
\author{Daisuke Nitta}
\affiliation{Department of Physics and Astrophysics, Nagoya University,
Aichi 464-8602, Japan}
\email{nitta@a.phys.nagoya-u.ac.jp}
\author{Shuichiro Yokoyama}
\affiliation{Department of Physics and Astrophysics, Nagoya University,
Aichi 464-8602, Japan}
\author{Kiyotomo Ichiki}
\affiliation{Department of Physics and Astrophysics, Nagoya University,
Aichi 464-8602, Japan}
\author{Keitaro Takahashi}
\affiliation{Graduate School of Science and Technology
Kumamoto University
2-39-1 Kurokami, Kumamoto 860-8555, Japan}

\date{\today}

\begin{abstract}

If the seed magnetic fields exist in the early Universe, tensor
 components of their anisotropic stresses are not compensated prior to
 neutrino decoupling and the tensor metric perturbations generated from
 them survive passively. Consequently, due to the decay of
 these metric perturbations after recombination, the so-called
 integrated Sachs-Wolfe effect, the large-scale fluctuations of CMB radiation are significantly
 boosted. This kind of CMB anisotropy is called the ``tensor passive mode.'' Because these fluctuations deviate largely from the Gaussian
 statistics due to the quadratic dependence on the strength of
 the Gaussian magnetic field, not only the power spectrum but also the
 higher-order correlations have reasonable signals. 
With these motives, we compute the CMB bispectrum induced by this mode.
When the magnetic spectrum obeys a nearly scale-invariant shape, we obtain an estimation of a typical value of the normalized reduced
 bispectrum as $\ell_1(\ell_1 + 1)\ell_3(\ell_3+1)|b_{\ell_1\ell_2\ell_3}| \sim (130-6) \times 10^{-16} ( B_{1 \rm Mpc} / 4.7 {\rm nG})^6$
depending on the energy scale of the magnetic field production from $10^{14}$GeV to $10^3$GeV.
Here, $B_{1 {\rm Mpc}}$ is the strength of the primordial magnetic field
 smoothed on $1 {\rm Mpc}$. From the above estimation and the current observational constraint on the primordial non-Gaussianity, we get a rough constraint on the magnetic field strength as $B_{1 {\rm Mpc}} < 2.6 - 4.4 {\rm nG}$.

\end{abstract}

\maketitle


\section{Introduction}\label{sec:intro}

Cosmological observations have suggested the existence
of micro-Gauss strength magnetic fields in galaxies and clusters of galaxies at the present Universe. As their origin, many researchers have discussed the possibility of generating the seed fields in the early Universe (e.g. \cite{Martin:2007ue, Bamba:2006ga}). 
These scenarios have been verified by constraining the strength of
the primordial magnetic fields (PMFs) through the effect on CMB fluctuations.

Conventional studies have provided upper bounds on PMFs with the two point
correlations (power spectra) of the CMB temperature and polarization anisotropies \cite{Paoletti:2010rx, Shaw:2010ea}.
On the other hand, taking into account the CMB three-point correlations
(bispectra), which have a nonzero value because the CMB fluctuations are sourced from the quadratic (non-Gaussian) terms of the stochastic (Gaussian) PMFs, some new consequences have been obtained. In
Refs.~\cite{Seshadri:2009sy, Caprini:2009vk, Trivedi:2010gi}, the
authors evaluated the contribution of the scalar modes at large
scale with several approximations, such as the Sachs-Wolfe limit, and roughly
estimated the upper limit on the PMF strength. In our
previous papers \cite{Shiraishi:2010yk, Shiraishi:2011fi}, we computed
the effect of the vector modes without neglecting the complicated angular
dependence, and obtained tighter bounds due to the dominant contribution
at small scale induced by the Doppler and the integrated Sachs-Wolfe (ISW) effects \cite{Mack:2001gc, Kahniashvili:2010us}. 
However, if the gravitational waves are generated from the PMF
anisotropic stresses uncompensated prior to neutrino decoupling, 
these superhorizon modes survive passively and the decay of their
modes after recombination amplifies the CMB anisotropies through the ISW effect \cite{Lewis:2004ef}. 
This type of fluctuation is called the ``tensor passive
mode'' and it is expected that the CMB bispectrum of this mode has the
most dominant signal at large scales, as inferred from the power spectrum \cite{Shaw:2009nf}.
Therefore, in this paper, we investigate the exact CMB bispectrum of
tensor passive modes induced from PMFs and place a new constraint on the
strength of PMFs. In the calculation, because there are complicated
angular integrals as there are in the vector mode case, we apply our
computation approach, as discussed in Ref.~\cite{Shiraishi:2011fi}.

This paper is organized as follows. In the next section, we formulate the CMB bispectrum of tensor passive modes induced from PMFs. In Sec. \ref{sec:result}, we show our result for the CMB bispectrum and the limit on the strength of PMFs, and give a discussion.

\section{Formulation of tensor bispectrum induced from PMFs}
\label{sec:formula}

Let us consider the stochastic PMFs $B^b({\bf x}, \tau)$ 
on the Friedmann-Robertson-Walker and small perturbative metric as 
\begin{eqnarray}
ds^2 = a(\tau)^2[- d\tau^2 + 2h_{0b} d\tau dx^b + (\delta_{bc} + h_{bc}) dx^b dx^c]
~.
\end{eqnarray}
Here $a$ is a scale factor and $\tau$ is a conformal time. 
In this space-time, the PMF evolves as $B^b({\bf x}, \tau) = B^b({\bf x}) /
a^2$. Then the spatial components of the PMF's energy momentum tensors are given by
\begin{eqnarray}
\begin{split}
T^b_{~c} &= \frac{1}{4\pi a^4} \left[\frac{B^2({\bf x})}{2}\delta^b_{~c} -  B^b({\bf x})B_c({\bf x})\right] \\
&\equiv \rho_\gamma \left(\Delta_B \delta^b_{~c} + \Pi^b_{Bc} \right)~, \label{eq:EMT_PMF} 
\end{split}
\end{eqnarray}
where 
$B^2 = B^b B_b$ and we use the photon energy density $\rho_{\gamma} (\propto a^{-4})$ for normalization.
In the following discussion, 
the index is lowered by $\delta_{bc}$, and the summation is implied for repeated indices.

\subsection{Bispectrum of the tensor anisotropic stress fluctuations}

The Fourier component of $\Pi_{B ab}$ is given by the convolution of the PMFs as
\begin{eqnarray}
\Pi_{B ab}({\bf k}) = -\frac{1}{4\pi \rho_{\gamma,0}}
\int \frac{d^3 {\bf k}'}{(2\pi)^3} B_a({\bf k}') B_b ({\bf k} - {\bf k}')~,
\label{eq:EMT}
\end{eqnarray}
where $\rho_{\gamma, 0} \equiv \rho_\gamma a^4$ denotes the present energy density of photons.
If $B^a({\bf x})$ obeys the Gaussian statistics,
the power spectrum of the PMFs $P_B(k)$ is defined by
\begin{eqnarray}
\Braket{B_a({\bf k})B_b({\bf p})}
= (2\pi)^3 {P_B(k) \over 2} P_{ab}(\hat{\bf k}) \delta ({\bf k} + {\bf
p})~, \label{eq:def_power}
\end{eqnarray}  
with a projection tensor
\begin{eqnarray} 
P_{ab}(\hat{\bf k}) 
\equiv \sum_{\sigma = \pm 1} \epsilon^{(\sigma)}_a \epsilon^{(-\sigma)}_b 
 =  \delta_{ab} - \hat{k}_a \hat{k}_b ~,
\end{eqnarray}
which comes from the divergenceless of the PMF.
Here $\hat{\bf k}$ denotes a unit vector, $\epsilon_a^{(\pm 1)}$ is a normalized divergenceless polarization vector satisfying the orthogonal condition, $\hat{k}^a \epsilon_a^{(\pm 1)} = 0$, and $\sigma (= \pm 1)$ expresses the helicity of the polarization vector. 
In general, the magnetic power spectrum should contain an
asymmetric helical term \cite{Caprini:2003vc, Kahniashvili:2005xe,
Pogosian:2001np}. However, we assume the magnetic fields are isotropic
and homogeneous, for simplicity; hence, this effect is neglected in Eq.~(\ref{eq:def_power}). Because the production mechanism of PMFs remains to be done, we use a simple power-law form as the power spectrum: 
\begin{eqnarray}
P_B(k) = {(2\pi)^{n_B+5} \over \Gamma(n_B / 2 + 3/2)k_{1 {\rm Mpc}}^3}
B_{1 {\rm Mpc}}^2 \left({k \over k_{1 {\rm Mpc}}} \right)^{n_B}~,
\end{eqnarray}
where $B_{1 {\rm Mpc}}$ denotes the magnetic field strength smoothed on
a scale $1 {\rm Mpc}$, $k_{\rm 1 Mpc} \equiv 2 \pi {\rm Mpc}^{-1}$, and $n_B$ is a spectral index.

With a transverse and traceless polarization tensor \cite{Shiraishi:2010kd}, 
$e_{ab}^{({\pm 2})} \equiv \sqrt{2} \epsilon_{a}^{(\pm 1)} \epsilon_{b}^{(\pm 1)}$, 
the anisotropic stress fluctuation is decomposed into two helicity states of the  tensor mode as 
\begin{eqnarray}
\Pi_{B ab}({\bf k}) = \sum_{\lambda = \pm 2} \Pi_{Bt}^{({\lambda})}({\bf k}) e_{ab}^{({\lambda})}(\hat{\bf k})~,
\end{eqnarray}
which is inversely converted into 
\begin{eqnarray}
\Pi_{Bt}^{({\pm 2})}({\bf k}) = \frac{1}{2} e_{ab}^{({\mp
 2})}(\hat{\bf k}) \Pi_{B ab}({\bf k})~.
\label{eq:tensani}
\end{eqnarray}

From the above equations, the bispectrum of $\Pi_{Bt}^{(\pm 2)}$ is symmetrically formed as
\begin{widetext}
\begin{eqnarray}
\Braket{\prod_{n=1}^{3} \Pi_{Bt}^{(\lambda_n)}({\bf k_n})}
&=& \left(-16 \pi \rho_{\gamma, 0}\right)^{-3} 
\left[ \prod_{n=1}^3 \int d^3 {\bf k_n'} P_B(k'_n) \right]
\delta({\bf k_1} - {\bf k_1'} + {\bf k_3'}) 
\delta({\bf k_2} - {\bf k_2'} + {\bf k_1'}) 
\delta({\bf k_3} - {\bf k_3'} + {\bf k_2'})
\nonumber \\
&& \times e^{(- \lambda_1)}_{ab} (\hat{\bf k_1}) 
e^{(- \lambda_2)}_{cd} (\hat{\bf k_2}) e^{(- \lambda_3)}_{ef} (\hat{\bf k_3})
[P_{ad}(\hat{\bf k_1'}) P_{be}(\hat{\bf k_3'}) P_{cf}(\hat{\bf k_2'}) + \{a \leftrightarrow b \ {\rm or} \ c \leftrightarrow d \ {\rm or} \ e \leftrightarrow f\}],
\label{eq:3-point}
\end{eqnarray}
\end{widetext}
where $\lambda$ means two helicities: $\lambda_1, \lambda_2, \lambda_3 = \pm 2$, and the curly brackets denote the symmetric 7 terms under the permutations of indices: $a \leftrightarrow b$, $c \leftrightarrow d$, or $e \leftrightarrow f$. 
Because the anisotropic stress fluctuation depends quadratically on the Gaussian magnetic fields as shown in Eq.~(\ref{eq:EMT}), the statistics of their tensor modes given by (\ref{eq:tensani}) is highly non-Gaussian. Hence, the bispectrum of Eq.~(\ref{eq:3-point}) also has a nonzero value and induces the finite CMB bispectrum. 

\subsection{CMB temperature bispectrum of tensor passive  modes}

As is well known, the gravitational potential of tensor modes can be
generated from anisotropic stresses via the Einstein equation.
If PMFs exist, the anisotropic
stresses, as mentioned in the previous subsection, also behave as a source. 
In general, after neutrino decoupling, the anisotropic stresses of
PMFs vanish via the compensation of those of neutrinos. However, prior to this epoch, there is no
compensation process due to the absence of the neutrino anisotropic
stresses. Hence, 
from the Einstein equation, we find the evolution equation of the tensor-mode metric perturbations as
\begin{widetext}
\begin{eqnarray} 
{h^{(\pm 2)}}''({\bf  k}, \tau) + 2 \frac{a'}{a} {h^{(\pm 2)}}'({\bf  k}, \tau)  + k^2 h^{(\pm 2)}({\bf  k}, \tau) 
\approx \left\{ \begin{array}{ll}
16 \pi G a^2 \rho_{\gamma} \Pi^{(\pm 2)}_{Bt}({\bf  k})  & (\tau_B \lesssim \tau \lesssim \tau_\nu) \\
0 & (\tau \gtrsim \tau_\nu) 
\end{array} \right. 
~, \label{eq:einstein}
\end{eqnarray}
\end{widetext} 
where $\tau_\nu$ and $\tau_B$ are the conformal times at neutrino decoupling
and the generation of the PMF, respectively, and $'$ denotes a derivative of
conformal time. Here $h^{(\pm 2)}$ is given by 
\footnote{$h^{(\pm 2)}$ is equal to $2 H_{T}$ of
Refs.~\cite{Lewis:2004ef, Shaw:2009nf}.} 
\begin{eqnarray}
h^{({\pm 2})}({\bf k},\tau) = \frac{1}{2} e_{ab}^{({\mp
 2})}(\hat{\bf k}) h_{ab}({\bf k},\tau) ~.
\end{eqnarray} 
From Eq.~(\ref{eq:einstein}), we find a superhorizon solution of the
tensor metric perturbation as \cite{Lewis:2004ef, Shaw:2009nf}
 \begin{eqnarray}
h^{(\pm 2)}({\bf k}) \approx h^{(\pm 2)}({\bf k}, \tau_\nu )
\approx 6 R_\gamma 
\ln\left(\frac{\tau_\nu}{\tau_B}\right) 
\Pi_{B t}^{(\pm 2)}({\bf k})~, \label{eq:ini_h}
\end{eqnarray} 
where $R_\gamma \sim 0.6$ is the ratio by the energy density of photons
to all relativistic particles for $\tau < \tau_\nu$.

The CMB temperature fluctuation is expanded into spherical harmonics as
${\Delta T (\hat{\bf n}) \over T} = \sum_{\ell m} a_{\ell m} Y_{\ell
m}(\hat{\bf n})$.
The $a_{\ell m}$ sourced from the initial tensor perturbations
(\ref{eq:ini_h}) can be expressed as \cite{Shiraishi:2010kd}
\begin{eqnarray}
a_{\ell m} &=& (-i)^{\ell} \int {k^2 dk \over 2\pi^2} \mathcal{T}_{\ell}(k) 
\sum\limits_{\lambda = \pm 2} h^{(\lambda)}_{\ell m}(k)~,
 \label{eq:alm} \\
h_{\ell m}^{({\pm 2})}(k) &\equiv& \int d^2 \hat{\bf k} h^{({\pm 2})}({\bf k}) 
{}_{\mp 2}Y^*_{\ell m}(\hat{\bf k})~. \label{eq:expand_h}
\end{eqnarray}
where $\mathcal{T}_{\ell}(k)$ denotes the transfer function of
tensor modes.
Because the solution of the magnetic passive mode (\ref{eq:ini_h}), if any, would dominate the
tensor-mode perturbation, the evolution of tensor modes after their
creation is almost identical to the standard cosmological one without
anisotropic stress sources. Therefore, we can use the standard
cosmological tensor-mode transfer function \cite{Zaldarriaga:1996xe, Weinberg:2008, Shiraishi:2010sm}.  

The CMB angle-averaged bispectrum is given by 
\begin{eqnarray}
B_{\ell_1 \ell_2 \ell_3} &=& 
\sum_{m_1 m_2 m_3} 
\left(%
\begin{array}{ccc}
  \ell_1  & \ell_2    & \ell_3   \\
   m_1    & m_2       & m_3   \\
\end{array}%
\right)
\Braket{\prod_{n=1}^3 a_{\ell_n m_n}},
\label{eq:cmb_bis_form}
\end{eqnarray}
where the bracket denotes the Wigner-$3j$ symbol.

In order to calculate the bispectrum of $a_{\ell m}$ given by
Eq.~(\ref{eq:alm}), 
we rewrite all angular dependencies in Eq.~(\ref{eq:3-point}) in terms of the spin-weighted spherical harmonics with the notation as \cite{Shiraishi:2010kd}
\begin{eqnarray}
\begin{split}
\epsilon^{(\pm 1)}_a(\hat{\bf r}) 
&= \epsilon^{(\mp 1) *}_a (\hat{\bf r})
 = \mp \sum_m \alpha^m_a {}_{\pm 1}Y_{1 m}(\hat{\bf r}) ~, \\
 \alpha^m_a \alpha^{m'}_a &= \frac{4\pi}{3}(-1)^m \delta_{m, -m'}~.
\end{split}
\end{eqnarray}
We then express the angular integrals of the spin spherical harmonics
with the Wigner-$3j$ symbols, and sum up these Wigner-$3j$ symbols over
the azimuthal quantum numbers in the same manner as in Ref.~\cite{Shiraishi:2011fi}. Then, we obtain the final form of the bispectrum as
\begin{widetext}
\begin{eqnarray}
\begin{split}
B_{\ell_1 \ell_2 \ell_3}
&=
(- 4 \pi \rho_{\gamma,0})^{-3}
\left[ \prod\limits^3_{n=1}
(-i)^{\ell_n}
\int {k_n^2 dk_n \over 2\pi^2}
\mathcal{T}_{\ell_n}(k_n)
\sum\limits_{\lambda_n = \pm 2}
\int_0^{k_D} k_n'^2 dk_n' P_B(k_n') \right] \\ 
&\quad \times 
\sum_{L L' L''} \sum_{S, S', S'' = \pm 1} 
\left\{
  \begin{array}{ccc}
  \ell_1 & \ell_2 & \ell_3 \\
  L' & L'' & L 
  \end{array}
 \right\}
f^{S'' S \lambda_1}_{L'' L \ell_1}(k_3',k_1',k_1) f^{S S' \lambda_2}_{L L' \ell_2}(k_1',k_2',k_2)
f^{S' S'' \lambda_3}_{L' L'' \ell_3}(k_2',k_3',k_3), \\ 
f^{S'' S \lambda}_{L'' L \ell}(r_3, r_2, r_1) 
&\equiv - 4 (8\pi)^{3/2} R_\gamma \ln \left( \tau_\nu \over \tau_B \right)
\sum_{L_1 L_2 L_3} \int_0^\infty y^2 dy j_{L_3}(r_3 y) j_{L_2}(r_2 y) j_{L_1}(r_1 y) \\
&\quad \times
(-1)^{\ell + L_2+L_3} 
(-1)^{(L_1 + L_2 + L_3)/2}
I^{0 \ 0 \ 0}_{L_1 L_2 L_3} 
I^{0 S'' -S''}_{L_3 1 L''} I^{0 S -S}_{L_2 1 L} 
I_{L_1 \ell 2}^{0 \lambda -\lambda} 
 \left\{
  \begin{array}{ccc}
  L'' & L & \ell \\
  L_3 & L_2 & L_1 \\
  1 & 1 & 2
  \end{array}
 \right\}~, \label{eq:cmb_bis}
\end{split}
\end{eqnarray}
\end{widetext}
where $j_l(x)$ is the spherical Bessel function, $k_D$ is the Alfv\'en-wave damping length scale, the $2
\times 3$ and $3 \times 3$ matrices in the curly brackets denote the
Wigner-$6j$ and $9j$ symbols, respectively, and 
\begin{eqnarray}
I^{s_1 s_2 s_3}_{l_1 l_2 l_3} 
\equiv \sqrt{\frac{(2 l_1 + 1)(2 l_2 + 1)(2 l_3 + 1)}{4 \pi}}
\left(
  \begin{array}{ccc}
  l_1 & l_2 & l_3 \\
   s_1 & s_2 & s_3 
  \end{array}
 \right)~. \nonumber 
\end{eqnarray}
As shown in Eq.~(\ref{eq:cmb_bis}), the bispectrum depends on
$\tau_B$. Although the production mechanism of PMFs is unclear and still
being discussed, we assume that PMFs arise sometime between the energy
scale of any grand unification theory and the electroweak
transition. Hence, in the computation of the CMB bispectrum, we consider
two corresponding values: $\tau_\nu / \tau_B \sim 10^{17}, 10^6$. This
leads to a factor of $23$ difference in the amplitude of the CMB
bispectrum due to the logarithmic dependence on $\tau_\nu /
\tau_B$. Therefore, due to the sextuplicate dependence of the CMB
bispectrum on the magnetic strength, there is a model-dependent
factor $23^{1/6} \simeq 1.7$ in bounds with the PMF strength.

\section{Numerical results and discussion}\label{sec:result}
Following the final expression (\ref{eq:cmb_bis}), we compute the CMB
temperature bispectrum of tensor passive modes numerically 
\footnote{Unlike the case of the vector mode bispectrum calculation \cite{Shiraishi:2011fi}, we do not use the thin LSS approximation because the temperature anisotropies from tensor modes are nonlocal.  
To check our numerical calculation, we computed the CMB power spectrum
of the tensor magnetic passive mode using the same method described in the main
text, namely, by expanding the nonlinear convolution of magnetic
anisotropic stress with the spin-weighted spherical harmonics. We
observe that our results are consistent with the previous results \cite{Shaw:2009nf}.}.

\begin{figure}[h]
  \begin{center}
    \includegraphics{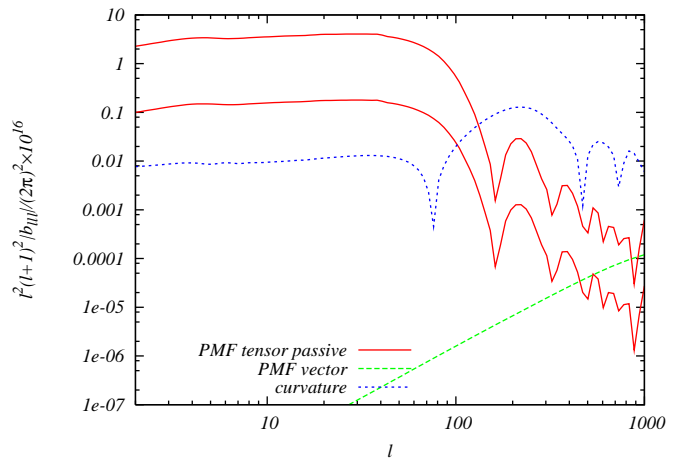}
  \end{center}
  \caption{(color online). Absolute values of the normalized reduced CMB bispectra given by Eq.~(\ref{eq:cmb_bis}) for $\ell_1 = \ell_2 = \ell_3$. 
The lines correspond to
 the spectra of tensor passive modes (red solid lines),
 vector modes \cite{Shiraishi:2010yk} (green dashed line), and primordial
 non-Gaussianity with $f^{\rm local}_{\rm NL} =
 5$ \cite{Komatsu:2001rj} (blue dotted line). The PMF parameters are
 fixed to $B_{1 {\rm Mpc}} = 4.7 {\rm nG}, n_B = -2.9$, and
 $\tau_{\nu}/\tau_{B} = 10^{17}$ (upper line) and $10^{6}$ (lower line), and the other cosmological
 parameters are equal to the mean values limited from the Wilkinson Microwave Anisotropy Probe 7-yr data \cite{Komatsu:2010fb}.}
  \label{fig:tens_pas_III_samel.eps}
\end{figure}

In Fig.~\ref{fig:tens_pas_III_samel.eps},
we describe the reduced bispectra of temperature fluctuations induced by the PMFs defined as~\cite{Komatsu:2001rj}
$b_{\ell_1 \ell_2 \ell_3}
\equiv 
 (I^{0 \ 0 \ 0}_{\ell_1 \ell_2 \ell_3})^{-1} B_{\ell_1 \ell_2 \ell_3}$,
for $\ell_1 = \ell_2 = \ell_3$. 
From the red solid lines, 
we can find that the enhancement at $\ell \lesssim 100$ due to the ISW
effect gives the dominant signal like in the
angular power spectrum $C_\ell$ \cite{Shaw:2009nf, Pritchard:2004qp}. 
The amplitude $\ell \sim {\cal O}(1)$ is comparable to $C_\ell^{3/2}$
because the power-law suppression of the Wigner symbols like the vector mode
\cite{Shiraishi:2010yk} is not effective at small $\ell$.

\begin{figure}[h]
  \begin{center}
    \includegraphics{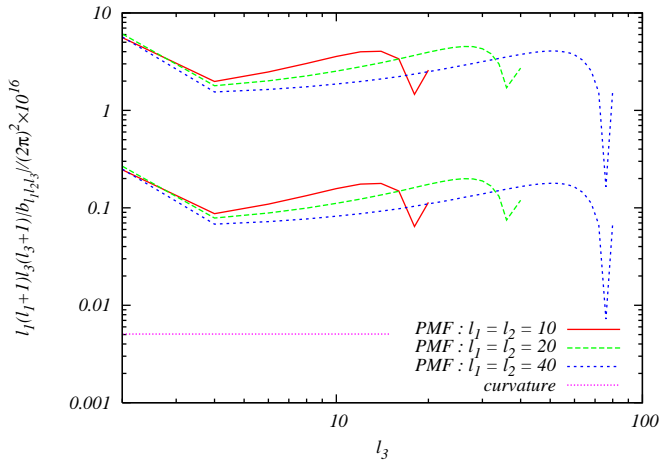}
  \end{center}
  \caption{(color online). Absolute values of the normalized reduced CMB
 bispectra given by Eq.~(\ref{eq:cmb_bis}) and
 generated from primordial non-Gaussianity in curvature perturbations given by
 Eq.~(\ref{eq:CMB_bis_local}) as a function of $\ell_3$ with $\ell_1 = \ell_2$.
Each parameter is identical to the values defined in Fig. \ref{fig:tens_pas_III_samel.eps}.
}
  \label{fig:tens_pas_III_difl.eps}
\end{figure}

In Fig.~\ref{fig:tens_pas_III_difl.eps}, we also show $b_{\ell_1 \ell_2
\ell_3}$ with respect to $\ell_3$ for $\ell_1 = \ell_2$. From this figure, for $n_B = -2.9$, the normalized reduced bispectrum is evaluated as 
\begin{eqnarray}
&& 
\ell_1 (\ell_1+1) \ell_3 (\ell_3+1) |b_{\ell_1 \ell_2 \ell_3}| 
\nonumber \\
&& \qquad 
\sim (130 - 6) \times 10^{-16} 
\left(\frac{B_{1 \rm Mpc}}{4.7 \rm nG} \right)^6~, 
\end{eqnarray}
where the factor $130$ corresponds to the $\tau_\nu / \tau_B = 10^{17}$
case and $6$ corresponds to $10^6$. It is also clear that $b_{\ell_1
\ell_2 \ell_3}$ for $n_B \sim -3$ dominates in $\ell_1 = \ell_2 \gg \ell_3$. 
Comparing this with the approximate expression of the bispectrum of local-type
primordial non-Gaussianity in curvature perturbations as \cite{Riot:2008ng}
\begin{eqnarray}
\ell_1 (\ell_1+1) \ell_3 (\ell_3+1)b_{\ell_1 \ell_2 \ell_3} \sim 4
 \times 10^{-18} f^{\rm local}_{\rm NL}~, \label{eq:CMB_bis_local}
\end{eqnarray}
the relation between the magnitudes of the PMF and the nonlinearity
parameter of the local-type configuration $f^{\rm local}_{\rm NL}$ is derived as
\begin{eqnarray}
\left( \frac{B_{1\rm Mpc}}{1\rm nG} \right) \sim (1.22 - 2.04)
|f^{{\rm local}}_{\rm NL}|^{1/6}~.
\end{eqnarray} 

Using the above equation, we can obtain the upper bound on the PMF strength.
As shown in Fig.~\ref{fig:tens_pas_III_samel.eps}, because the tensor bispectrum is highly damped for $\ell \gtrsim 100$, 
we should use an upper bound on $f^{\rm local}_{\rm NL}$ obtained by the current observational data for $\ell < 100$, namely $f^{\rm local}_{\rm NL} < 100$ \cite{Smith:2009jr}. This value is consistent with a simple prediction from the cosmic variance \cite{Komatsu:2001rj}. 
From this value, we derive $B_{1 \rm Mpc} < 2.6 - 4.4 {\rm nG}$
These are $4-2$ times stronger than vector-modes bounds
\cite{Shiraishi:2010yk}.
 
In this paper, we study the CMB temperature bispectrum generated from
the tensor anisotropic stresses of PMFs and find a new constraint on the magnetic field magnitude when the PMF spectrum is close to a scale-invariant shape. Although there is a touch of uncertainty in the production epoch of PMFs, 
this bound is tighter than ones obtained by the CMB power spectra
\cite{Paoletti:2010rx, Shaw:2010ea}. 
Although this limit is weaker than a rough bound from only the
scalar passive modes \cite{Trivedi:2010gi} due to the rapid decay of the tensor bispectrum at small scales, 
the significant amplitude at large scales will have a drastic impact on
the precise calculation of the limit on PMFs, including the scalar, vector, and
tensor-mode contributions.

In our previous studies and the above analysis, we find that tensor (vector) modes dominate at
large (small) scale, not only in the power spectrum but also in the
bispectrum. It is also expected that the scalar mode dominates at the intermediate scale.
Therefore, using this scale-dependent property, we will also constrain a
spectral index of the PMF spectrum in addition to the magnetic strength. These reasonable bounds will be obtained by considering the CMB temperature and polarization bispectrum of autocorrelations and cross-correlations between scalar, vector, and tensor modes in the estimation of the signal-to-noise ratio.

\begin{acknowledgments}
We would like to thank Dai G. Yamazaki for useful discussions.
This work is supported by the Grant-in-Aid for JSPS Research under
 Grant No.~22-7477 (M. S.), and JSPS Grant-in-Aid for Scientific
Research under Grants No.~22340056 (S. Y.), No.~21740177, No.~22012004 (K. I.),
 and No.~21840028 (K. T.).
This work is supported in part by the Grant-in-Aid for Scientific
Research on Priority Areas No. 467 "Probing the Dark Energy through an
 Extremely Wide and Deep Survey with Subaru Telescope" and by the
 Grant-in-Aid for Nagoya University Global COE Program "Quest for
 Fundamental Principles in the Universe: from Particles to the Solar
 System and the Cosmos," from the Ministry of Education, Culture,
 Sports, Science and Technology of Japan. 
\end{acknowledgments}

\bibliography{paper}
\end{document}